\documentclass[rmp,aps,preprint]{revtex4-1}

\usepackage{graphicx, graphics}

\usepackage{blindtext,tikz}
\usetikzlibrary{calc}
\usepackage{datetime}

\usepackage{amsthm, amsmath, amssymb}
\usepackage{geometry}

\begin{document}

\title{A Statistical Modelling Approach to Detecting Community in Networks}

\author{A. A. Ickowicz}
\affiliation{CSIRO Computational Informatics, North Ryde 1670 NSW, Australia}

\begin{abstract}

There has been considerable recent interest in algorithms for finding communities in networks - groups of vertex within which connections are dense (frequent), but between which connections are sparser (rare).   Most of the current literature advocates an heuristic approach to the removal of the edges (i.e., removing the links that are less significant using a well-designed function). In this article, we will investigate a technique for uncovering latent communities using a new modelling approach, based on how information spread within a network. It will  prove to be easy to use, robust and scalable. It makes supplementary information related to the network/community structure (different communications, consecutive observations) easier to integrate. We will demonstrate the efficiency of our approach by providing some illustrating real-world applications, like the famous Zachary karate club, or the Amazon political books buyers network.

\end{abstract}

\maketitle


\section{Introduction}
\label{sec:intro}
\begin{figure}
\includegraphics[trim=0cm 1cm 0cm 2cm, clip=true, scale = 0.45]{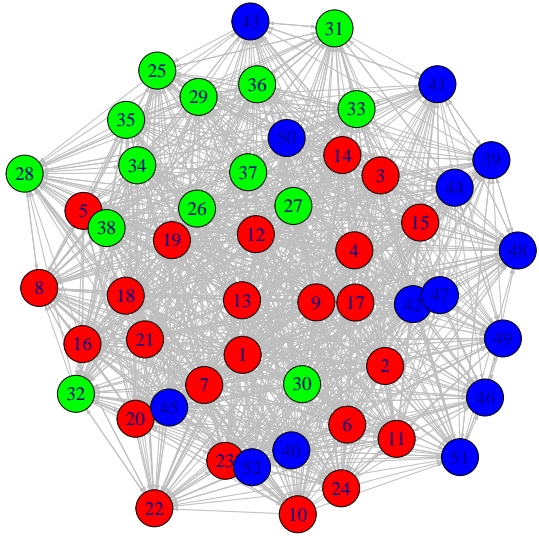}
\caption{\label{fig:net} Example of network.}
\end{figure}

In the continuing flurry of research activity within physics and mathematics on the properties of networks, a particular recent focus has been the analysis of communities within networks. In the simplest case, a network or graph can be represented as a set of points, or vertex, joined in pairs by lines, or edges. Many networks, it is found, are inhomogeneous, consisting not of an undifferentiated mass of vertices, but of distinct groups. Within these groups there are many edges between vertex, but between groups there are fewer edges, producing a structure like that sketched in Fig. \ref{fig:net}. The ability to find communities within large networks in some automated fashion could be of considerable use. Communities in a web graph for instance might correspond to sets of web sites dealing with related topics \citep{Adamic2005}, while communities within a Karate club are simply kids going to the same schools see \citep{Zachary1977}.\\
In many settings where we observe networks of interactions there are natural groupings of nodes so that pairs of nodes that are in the same group tend to interact more than pairs of nodes that belong to different groups. If nodes are people, they may belong to the same club, be of the same ethnicity or profession. In the case of trade unions, for example, individuals with similar jobs are more likely to interact like scientific researchers having collaborations and publishing together \citep{Newman2001}. This tendency of individuals to have a tendency to interact with others who have similar characteristics is called homophily, and is quite pervasive. In many cases, however, the underlying structure that influences network interactions is of interest but is not directly observable. In such cases we can only infer which nodes should be grouped together by observing their interaction patterns. 

\subsection{Heuristic Approaches}

A first class of algorithm to perform such an inference is mainly of heuristic nature. The \emph{Betweenness} algorithm proposed by \cite{Newman2003a} is a hierarchical decomposition process where edges are removed in the decreasing order of their edge betweenness scores (i.e. the number of shortest paths that pass through a given edge). This is motivated by the fact that edges connecting different groups are more likely to be contained in multiple shortest paths simply because in many cases they are the only option to go from one group to another. Another  hierarchical approach is proposed by \cite{Clauset2004} and is bottom-up instead of top-down when compared to the previously cited algorithm. It tries to optimize a quality function called modularity in a greedy manner. Initially, every vertex belongs to a separate community, and communities are merged iteratively such that each merge is locally optimal (i.e. yields the largest increase in the current value of modularity). \cite{Newman2006} proposed a top-down hierarchical approach still based on the modularity function but this time introducing the eigenvectors. In each step, the graph is split into two parts in a way that the separation itself yields a significant increase in the modularity. The split is determined by evaluating the leading eigenvector of the so-called modularity matrix, and there is also a stopping condition which prevents tightly connected groups to be split further. Random walk has been used by \cite{Pons2005} and \cite{A.Tabrizi2013}. The general idea is that if you perform random walks on the graph, then the walks are more likely to stay within the same community because there are only a few edges that lead outside a given community. Pons' walktrap runs short random walks of $3$ to $5$ steps (generally) and uses the results of these random walks to merge separate communities in a bottom-up manner. The latter proposed a top-down algorithm mixing both random walk and the modularity function that can reveal inherent clusters of a graph more accurately than other nearly-linear approaches that are mainly bottom-up.\\
These methods yield generally good results but can suffer from computational complexity, sensibility to resolution limit, and no modelling assumption is made. Therefore, no convergence has been derived for them, which make their usage uncertain.

\subsection{Modelling Approaches}

In addition to all these works on heuristic approaches, modelling-based approach have also been considered by various authors \citep{Karrer2011, Bickel2013}, through the Latent variable models. This main model, known as the stochastic blockmodel, assume that the network connections are explainable by a latent discrete class variable associated with each node. For this model, consistency has been shown for profile likelihood maximization \citep{Bickel2009}, a spectral-clustering based method \citep{Rohe2011}, and other methods \citep[see for example][]{Chen2011a}, under varying assumptions on the sparsity of the network and the number of classes. Other works, including the Infinite Relational Model (IRM) and the Infinite Hidden Relational Model \citep{Kemp2006,Xu2006}, allow a potentially infinite number of clusters. The Mixed Membership Stochastic Block Model (MMSB) \citep{Airoldi2006} increases the expressiveness of the latent class models by allowing mixed membership, associating each object with a distribution over clusters. \cite{Palla2012} conveys the idea of using an infinite latent attribute  model.\\
These results suggest that the model has reasonable statistical properties, and empirical experiments suggest that efficient approximate methods may suffice to find the parameter estimates. Two main drawbacks can be identified in these approaches. First, the optimization procedure becomes more and more complex as the dimension of the latent space increase. Secondly, despite it claims on being a modelling approach, our view is that it only models half of the problem, the communities of vertex. No mention is made on how the vertex connect, i.e. how we end up with a specific adjacency matrix.

\subsection{Contribution}

In our paper, we investigate a technique for uncovering latent communities using a new modelling approach, addressing the two issues pointed. Unlike many authors considering the model at the vertex level, we will also consider a modelling approach for the information. Indeed, we will consider that the observed network is not only the realization of a random variable, but essentially the result of the spread of many information impulses across the network. Using a $1$-dimensional latent model for the vertex and the spreading model, we will infer the value of the latent variables using a MCMC. In order to speed up the MCMC process, we also propose a parallel tempering-alike strategy in section \ref{sec:Af}. Our approach is easy to scale to large networks and suffer less from the curse of dimensionality. Supplementary information related to the network (different communications, consecutive observations) is also easy to integrate.\\
The paper is organised as follows. First, we make a brief recall on the stochastic blockmodels, and we introduce the notations that we will use for the rest of the paper. Then we present our contribution in the modelling section, detailing the different major differences with the approaches considered in the literature so far. The estimation procedure is presented in section \ref{sec:estimation}. We extend our approach with an additional algorithm in section \ref{sec:Af}, and finally provide some illustrating applications with real-world data.

\section{Preliminaries}
\label{sec:pr}

\subsection{Stochastic blockmodel}

We consider a class of latent variable models which we describe as follows. Let $Z = (Z_1,\dots ,Z_n)$ be latent random variables corresponding to vertex $1,\dots,n$. In the original stochastic blockmodel, they are taking values in $[K] = \{1, . . . , K\}$. $\pi$ is usually defined as a distribution on $[K]$, and $H$ as a symmetric matrix in $[0, 1]^{K \times K}$ . The \emph{complete graph model} (CGM) for $Z$, $A$, where $A$ is the $n \times n$ symmetric $0-1$ adjacency matrix of a graph, is defined by its distribution,
\begin{eqnarray}
\label{eq:dist_adj}
f(Z,A) &=& \Big(\prod_i^{n} \pi(Z_i) \Big) \Big( \prod_i^n \prod_{j = i+1}^n H(Z_i, Z_j)^{A_{ij}} (1 - H(Z_i, Z_j))^{1-A_{ij}}\Big),
\end{eqnarray}
where we may interpret $H(Z_i,Z_j)$ as $p( \textrm{edge} | Z_i, Z_j )$, and $\pi(a)$ as $p(Z_i = a)$ for $a = 1,\dots, K$.\\
The \emph{graph model} (GM) is defined by a distribution $g : \{0, 1\}^{n\times n} \rightarrow [0, 1]$, which satisfies $g(A) = p(A; H, \pi)$ and is given by
\begin{eqnarray}
g(A) &=& \sum_{z \in [K]^n} f(z,A)
\end{eqnarray}

\subsection{More general models}

The stochastic blockmodel is a special case of a more general latent variable model (see \cite{Hoff2002}). In this model, the elements of $Z$ take values in a general
space $Z$ rather than $[K]$, $\pi$ is a distribution on $Z$, and $H$ is replaced by a
symmetric map $h : Z \times Z \rightarrow [0, 1]$. The CGM defines a density for $(Z, A)$, with respect to an appropriate reference measure, and GM satisfies the identity
\begin{eqnarray}
\frac{g(A; \theta)}{g_0(A)} &=& \mathbb{E}_{\theta_0} \Big[ \frac{f(Z,A; \theta)}{f_0(Z,A)} {\Big\vert} A \Big]
\end{eqnarray}

\section{Modelling}
\label{sec:modelling}

The contribution of our article is described in this section. 
In our work, we will focus on a one dimensional latent space to describe the community structure, which allows more simplicity and is then less computationally demanding. We will demonstrate that using an appropriate latent space and link function, we can perform as well as multiple dimensional models.

\subsection{Latent Variable}
Our idea is to consider that the latent variable $Z_i$ of vertex $i$ evolves in the continuous circular space $\Theta = [0, 2\pi]$. The latent variable $\theta_i$ will represent the location of the vertex $i$ in the latent space. We make the modelling assumption that two vertex in the same community will have close $\theta_{1:n}$'s on the latent space. Then, if we consider the previous modelling, we define,
\begin{eqnarray}
H(Z_i, Z_j) &=& h(\theta_i - \theta_j)
\end{eqnarray}
where $h()$ stands for a link function defined below. 

\paragraph{Link function $h()$}

The main constraint on $h$ is that it should map $[0, 2\pi]$ on an Euclidean space $E$. For simplicity, we have chosen $E = [0,1]$. Theoretically, we would like that two vertex with opposite $\theta$ values have a very small probability of being in the same community. Therefore, we considered the following function,
\begin{eqnarray}
h(y) &=& \frac{\cos(y)+1}{2}, \quad \forall y \in [0, 2\pi]
\end{eqnarray}

\subsection{Modelling the information spread}

In the recent literature, the cells of the adjacency matrix are essentially considered as the realization of random variables (or a function of random variables). In this paper we aim at introducing a more real-life related interpretation (and subsequently modelling) of the adjacency matrix $A$. We consider that an edge is built between two vertex if an information is shared between two vertex, e.g. during a phone call, or a meeting, etc. We will call this "share of information" an \emph{impulse}. Depending on the kind of information, the impulse does not necessarily involve two vertex from the same community. However, we make the assumption that most of the time, it does\footnote{And it doesn't look like an unusual assumption on a statistical point of view.}. In this article, we define two kind of impulses. The \emph{sequential impulses}, where an information is spread from one vertex to another sequentially (like successive phone calls). And the \emph{instantaneous impulses}, where the information is spread to several vertex at the same time (meeting, diners...).\\
Let $\mathcal{I}$ be a shared information (or impulse), $\mathcal{I}$ is a $n \times n$ matrix which elements belong to $\{0,1\}$,
\begin{equation}
\mathcal{I}_{ij} = 
\left\{
\begin{array}{ll}
1 \textrm{ if vertex $i$ and $j$ exchange an information,}\\
0 \textrm{ otherwise.} 
\end{array}
\right.
\end{equation}
We assume that the adjacency matrix $A$ is observed after the exchange of several impulses.
\paragraph{Sequential impulse}

If we denote $N$ the number of information, and $T_i$ their respective propagation length,
\begin{eqnarray}
A_{\mathcal{I}} &=& \sum_i^N \sum_t^{T_{i}} \mathcal{I}_{it}
\end{eqnarray} 
In the case of sequential impulse, $\mathcal{I}$ doesn't have to be symmetric, and in our mind is designed to model directed graphs. However, this is not compulsory and symmetric undirected graph may also be created.

\paragraph{Instantaneous Impulse}
In the case of instantaneous impulses, $\mathcal{I}$ is not restricted to only two vertex sharing information, but we make the assumption that all the vertex are given the information at the same time. Therefore,
\begin{eqnarray}
A_{\mathcal{I}} &=& \sum_i^N \mathcal{I}_{i}
\end{eqnarray} 
These impulses can also be considered as community-based impulses and thus identifying the propagation of the information helps in identifying the communities. It is also worth notice that the kind of impulse suitable to the modelling is related to the data that you have.\\
Finally, depending on the output type, 
\begin{equation}
A = 
\left\{
\begin{array}{ll}
A_{\mathcal{I}} & \textrm{if } A \in \mathbb{N}^{n \times n}\\
\min(A_{\mathcal{I}},1) & \textrm{if } A \in \{0,1\}^{n \times n}
\end{array}
\right.
\end{equation}

\section{Estimation}
\label{sec:estimation}

Our latent model allows for some dimension reduction, but for large networks, the number of parameters ($\theta$) to be estimated is still very important, and then unlikely to be done through any optimization method in a reasonable amount of time. For small networks, the Bayesian paradigm can be used, and the posterior distribution of the $\theta$'s can be expressed as,
\begin{eqnarray}
\label{eq:post}
\pi(\theta_{1:n} \vert A, h) & \propto & p(A \vert \theta_{1:n}, h ) \times \pi_0(\theta_{1:n})
\end{eqnarray}
The use of an appropriate Monte Carlo simulation with an appropriately designed likelihood can then help us estimate the $\theta$'s.\\
For larger network though, the total run length needed will be too important for our technique to be a real breakthrough. Then, we decided to sub-sample the network using the modelling described in Section \ref{sec:modelling}, and apply the Bayesian approach to this sample. 

\subsection{Sub-sampling from the network using $A$}

The idea is that the natural spread of an information indicates the community links. Of course, the spread can also reach vertex that do not belong to the community, but \emph{most likely} they will. Then, if we simulate the spread of several informations $\mathcal{I}$, we should estimate their most relevant latent variables respectively to the other's. The steps are then the following,
\begin{itemize}
\item Simulate the spread of an information $\mathcal{I}$. Let $\sigma(i)$ be the list a attained vertex;
\item For each of the vertex in $\sigma(i)$, calculate $\theta$ using Eq. \ref{eq:post2}.
\end{itemize}
\begin{eqnarray}
\label{eq:post2}
\pi(\theta_{\sigma(i)} \vert A_{\sigma(i)}, h) & \propto & p(A_{\sigma(i)} \vert \theta_{\sigma(i)}, h ) \times \pi_0(\theta_{\sigma(i)})
\end{eqnarray}
where $A_{\sigma(i)}$ is the restriction of $A$ to the $\sigma(i)$'s vertex.

\paragraph{Number of simulated $\mathcal{I}$:} There is no upper limit to the number of $\mathcal{I}$ we can simulate. However, the only rule is as usual, "the more the better".

\paragraph{Size of each simulated $\mathcal{I}$:} This very much depends on the data that you have. Given that each $\mathcal{I}$ is supposed to spread along the community, the size should be big enough to allow $\mathcal{I}$ to spread to the entire community, but not to big so that it doesn't spread outside it.

\subsection{Calculate the likelihood}

In Eq. \ref{eq:post2}, the likelihood can be approximated according to the following equations, 
\begin{eqnarray}
p(A_{\sigma(i)} \vert \theta_{\sigma(i)}, h ) &=& \prod_{\substack{j,k \in \sigma(i)\\ j < k}} f(A_{\sigma(i)}^{j,k} \vert \theta_{\sigma(i)}, h)
\end{eqnarray}
Then, depending on $A$, $f$ is a probability density function that can have different form, in particular,
\begin{eqnarray}
f(A_{\sigma(i)}^{j,k} \vert \theta_{\sigma(i)}, h) &=& \mathcal{B}_{h(\theta_j - \theta_k)}(A_{\sigma(i)}^{j,k})
\end{eqnarray}
if $A \in \{0,1\}^{n \times n}$. We denote $\mathcal{B}_{\alpha}(x)$ the probability density function of the Bernoulli distribution at point $x$ with parameter $\alpha$. If  $A \in \mathbb{N}^{n \times n}$,
\begin{eqnarray}
f(A_{\sigma(i)}^{j,k} \vert \theta_{\sigma(i)}, h) &=& \mathcal{P}_{\lambda h(\theta_j - \theta_k)}({A_{\sigma(i)}^{j,k}})
\end{eqnarray}
where $\mathcal{P}_{\lambda}(x)$ the probability density function of the Poisson distribution at point $x$ with parameter $\lambda$.

\subsection{Evaluate the link between two vertex}

Once the posterior distribution of the $\theta_{\sigma(i)}$'s obtained, we want to integrate the different estimation together. The strength of this approach is that the distance between two vertex is easily determined by simply calculating the distance between the corresponding $\theta$'s. So, for two different information $\mathcal{I}$ and $\mathcal{I}'$, $\theta_{\sigma(i)}$ will not be equal to $\theta_{\sigma(i')}$, but the difference of the $\theta$'s should be. Then, the probability of "belonging to the same community" can be calculated using $h()$ and the estimation. For notational convenience, $\theta_{\sigma(i)}$ being a vector, we will denote $\theta_{\sigma(i)}[j]$ the $j$th element of $A$ that also belong to $\sigma(i)$. Then, 
\begin{eqnarray}
\hat{p}(j,k \in \mathcal{C} \vert j,k \in \sigma(i)) &=& h(\theta_{\sigma(i)}[j],\theta_{\sigma(i)}[k])
\end{eqnarray}
And for several information spread, 
\begin{eqnarray}
\hat{p}(j,k \in \mathcal{C} ) &=& \sum_i \hat{p}(j,k \in \mathcal{C} \vert j,k \in \sigma(i)) \hat{p}(j,k \in \sigma(i))
\end{eqnarray}
and $\hat{p}(j,k \in \sigma(i))$ can be approximated by,
\begin{eqnarray}
\hat{p}(j,k \in \sigma(i)) &=& \frac{1}{N} \sum_i {\bf 1}(j,k \in \sigma(i))
\end{eqnarray}
Then, the aggregation of successive simulated information spread can be done by simply summing the new estimated probabilities (and dividing by the number of estimates).

\subsection{Partitioning into communities}

Creating the partition can be done using the estimated probabilities. Using them, we build a Clustering Tree (or Dendrogram). This tree can then be pruned to identify the communities in the network.

\subsection{Selecting the best partition}

The algorithm induces a sequence $(\mathcal{P}_k)_{1 \leq k \leq N}$ of partitions into communities. We now want to know which partitions in this sequence capture well the community structure. We will use the modularity $Q$, a quality function widely used in recent community detection approaches \citep[see for example][]{Clauset2004, Pons2005, Francisco2011, Zhang2012} and introduced in \cite{Newman2003a, Newman2004}. It relies on the fraction of edges inside community and the fraction of edges bound to this community:
\begin{eqnarray}
Q &=& \textrm{tr}({\bf e}) - \| {\bf e}^2 \|
\end{eqnarray}
where $\textrm{tr}(M)$ stands for the trace of the matrix $M$, and ${\bf e}$ is the $k \times k$ matrix whose component $e_{ij}$ is the fraction of edges in the original network that connect vertices in group $i$ to vertices in group $j$. The best partition(s) is (are) then considered to be the one(s) that maximize(s) Q.

\section{Advanced features}
\label{sec:Af}

\subsection{Speeding up the MCMC run}

Even with a lower dimension problem, the MCMC run can last quite a long time before it reaches the stationary distribution. Our situation, though, present a specific feature that we can take advantage of. Indeed, some of the latent variables are likely to be very close if their vertex belong to the same community. Therefore, if we knew that two vertex were in the same community, we could make their latent variables "jump" to the same distribution, hence a gain of iteration of the MCMC run. This technique is inspired from the parallel tempering algorithm by \cite{Swendsen1986}, which allows parallel chains to jump to each other values with a certain probability. Our algorithm is the following,
\begin{itemize}
\item[$\small \mathbf{1:}$] Choose randomly one chain (or vertex), $c_i$;
\item[$\small \mathbf{2:}$] Choose randomly another chain $c_j$, with probability,
\begin{eqnarray}
p(c_j \vert c_i) &=& (\mathbf{1}_{\{A_{ij} = 1\}} + 1/n) / (\sum_j A_{ij} + 1)
\end{eqnarray}
\item[$\small \mathbf{3:}$] Accept the jump $\theta_i \leftarrow \theta_j$ with probability equal to the likelihood ratio of these two $\theta$'s.
\end{itemize}
The probability calculated in the second iteration is designed to select preferably vertex with links, but still allows non directly linked vertex to be chosen.

\subsection{Splitting Algorithm}

\begin{figure}
\centering
\includegraphics[trim=2cm 0cm 2cm 0cm, clip=true, scale = 0.23]{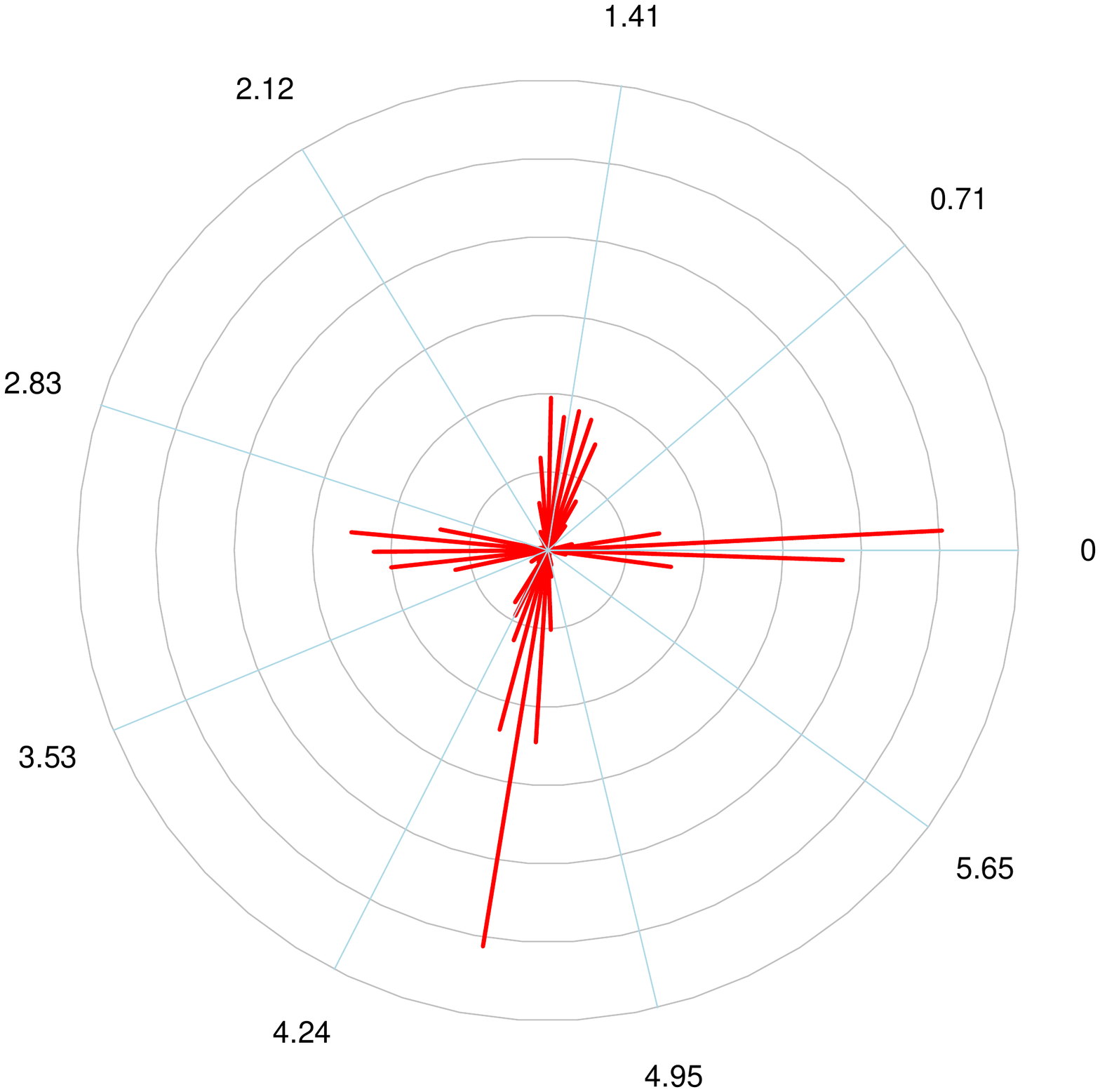}
\includegraphics[trim=2cm 0cm 2cm 0cm, clip=true, scale = 0.23]{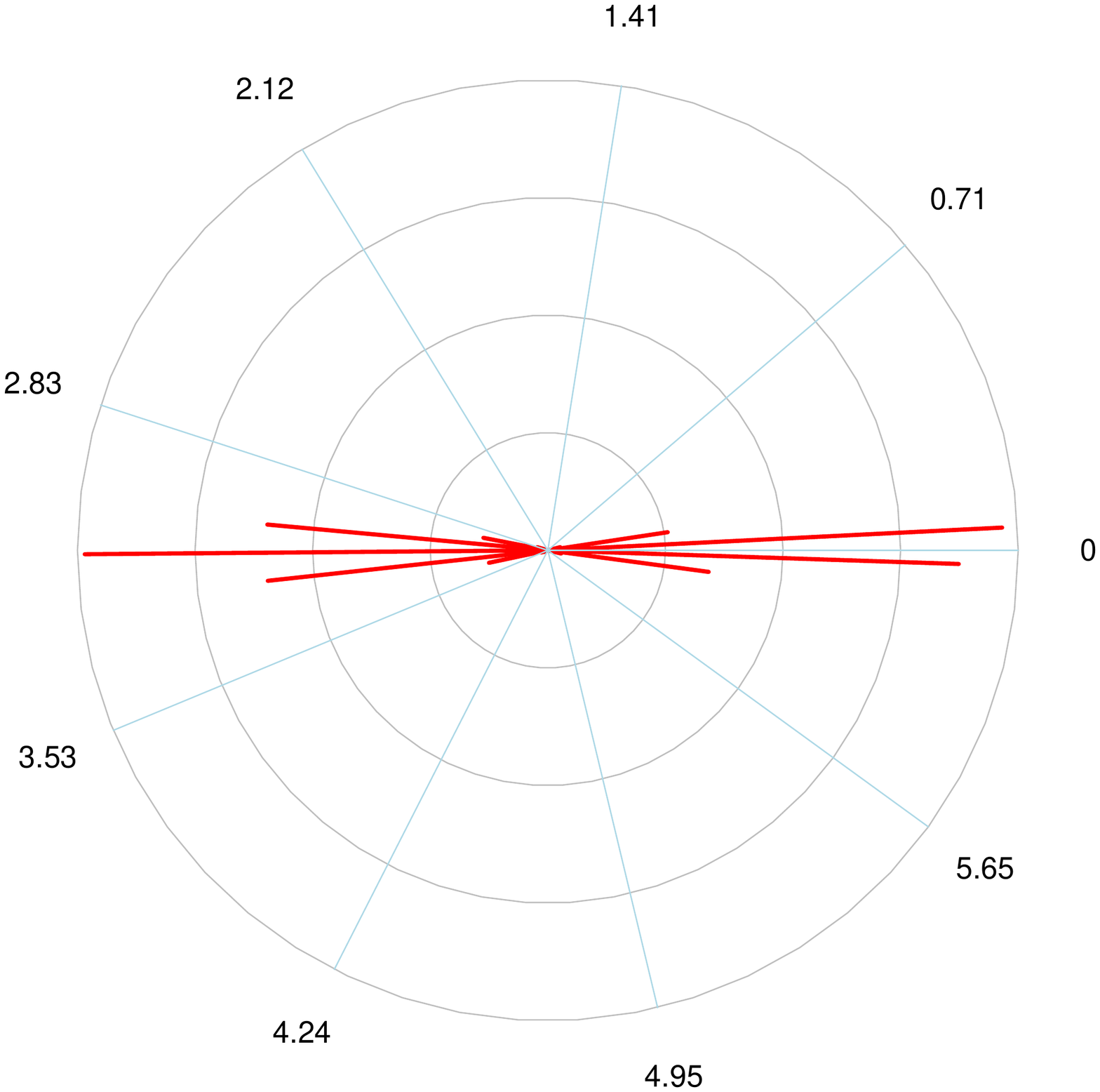}\\
\includegraphics[trim=2cm 0cm 2cm 0cm, clip=true, scale = 0.23]{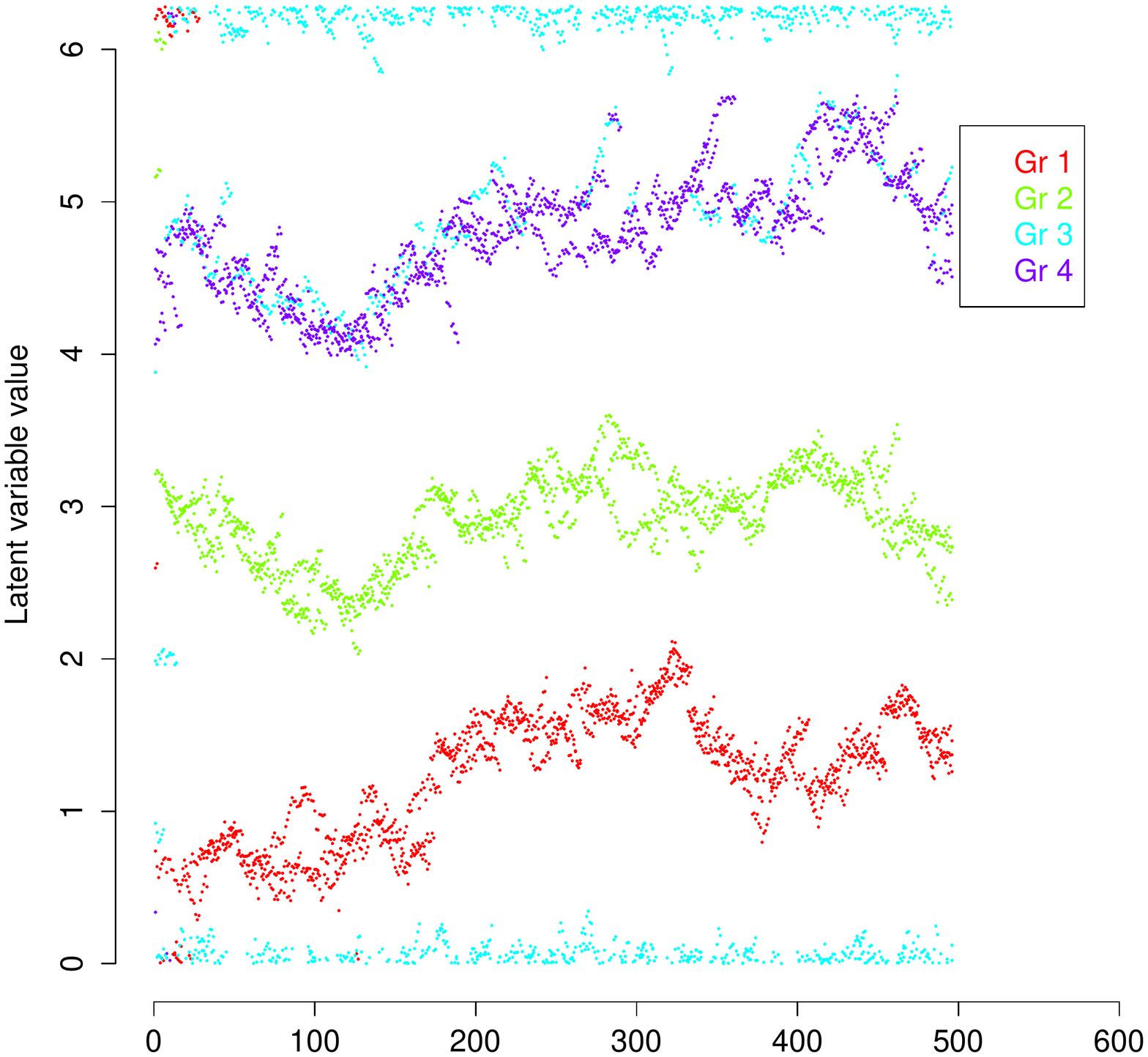}
\includegraphics[trim=2cm 0cm 2cm 0cm, clip=true, scale = 0.23]{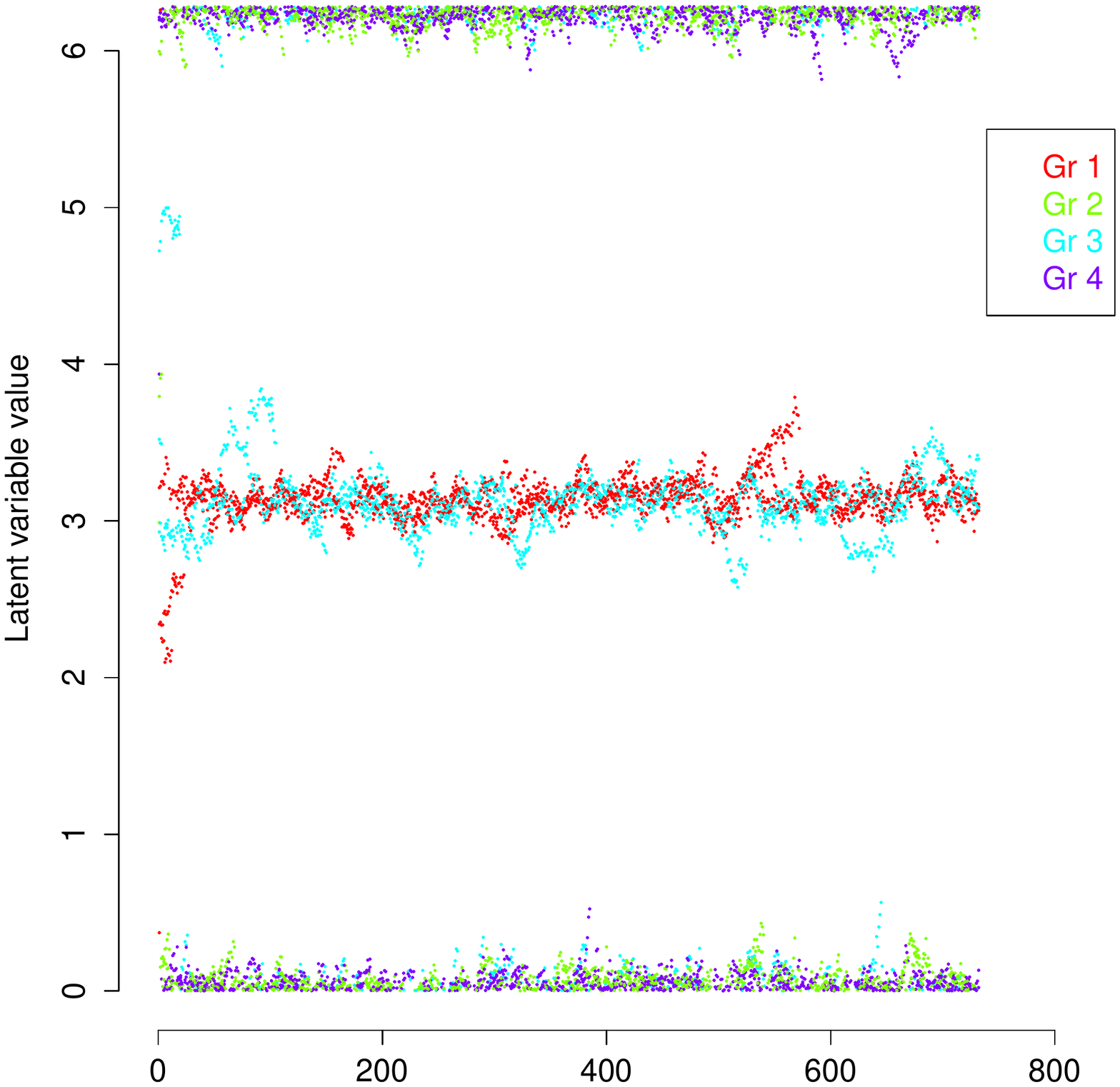}
\caption{\label{fig:BSvsMS} MCMC run and posterior distribution histogram after one MCMC run on a specific network. The left figures display the use of $h()$ in the likelihood, the figures on the right the use of $g()$. The number of iteration are the same, $3000$, with acceptance ratios of $0.17$ and $0.25$ respectively. The number of vertex is $12$, theoretically belonging to $4$ different communities.}
\end{figure}

The main drawback of this one-dimensional latent variable approach is the issue we have to face when the number of different communities is important. In that case, the run length has (also) to be very important to provide good estimating values of the parameters. Moreover, the difference between the latent variables may not be significant enough. So, the idea is to use the latent variables as clustering variables in order to undermine which vertex are likely to be in the same communities. Therefore, we aim at building a function that could help us separating iteratively the vertex in two groups. The general idea of this approach is known as iterative bisection, and is reputed as a Computer Science approach. We then decided to develop an equivalent algorithm using our framework.\\
The only difference with the previous approach is the link function. Instead of using $h()$, we will use $g()=h()^k, k\in \mathbb{N}$, which has a greater separating power. The algorithm is then modified as follows,
\begin{itemize}
\item[$\small \mathbf{1:}$] Simulate the spread of an information $\mathcal{I}$. Let $\sigma(i)$ be the list a attained vertex;
\item[$\small \mathbf{2:}$] Calculate $\theta$ using the new link function $g()$.
\item[$\small \mathbf{3:}$] Group the vertex using the estimated $\theta$.
\item[$\small \mathbf{4:}$] For each group, recalculate the $\theta$ using $g()$.
\item[$\small \mathbf{5:}$] Start over, until there is a maximum (previously set) number of vertex in each group.
\end{itemize}
We display in Fig \ref{fig:BSvsMS} the difference between the use of $h()$ and $g()$ in the MCMC step (Eq. \ref{eq:post2}).

\section{Results}

In the first two examples, we deal with real data where the results of our algorithm can be cross check with a ground truth. This comparison is performed using the Rand index corrected by \cite{Arabie1985} \citep[see also][]{Meila2005}. The Rand index $\mathcal{R}(\mathcal{P}_1,\mathcal{P}_2)$ is the ratio of pairs of vertices correlated by the partitions $\mathcal{P}_1$ and $\mathcal{P}_2$ (two vertices are correlated by the partitions $\mathcal{P}_1$ and $\mathcal{P}_2$ if they are classified in the same community or in different communities in the two partitions). The expected value of $\mathcal{R}$ for a random partition is not zero. To avoid this, the corrected index (which is also more sensitive) is introduced \citep[see][]{Vinh2009}: $\mathcal{R}'= \frac{\mathcal{R}-\mathcal{R}_{exp}}{\mathcal{R}_{max} - \mathcal{R}_{exp}} $ where $\mathcal{R}_{exp}$ is the expected value of $\mathcal{R}$ for two random partitions with the same community size as $\mathcal{P}_1$ and $\mathcal{P}_2$. This quantity has many advantages compared to the 'ratio of vertices correctly identified' that has been widely used in the past. It captures the similarities between partitions even if they do not have the same number of communities, which is crucial here as we will see below. Moreover, a random partition always gives the same expected value 0 that does not depend on the number of communities.

\subsection{Zachary's Karate Club}

This famous data considered the following real-world scenario. The karate club was observed for a period of three years, from $1970$ to $1972$. For its classes, the club employed a part-time karate instructor, who will be referred to as Mr. Hi. At the beginning of the study there was an incipient conflict between the club president, John A., and Mr. Hi over the price of karate lessons. Mr. Hi, who wished to raise prices, claimed the authority to set his own lesson fees, since he was the instructor. John A., who wished to stabilize prices, claimed the authority to set the lesson fees since he was the club's chief administrator. As time passed the entire club became divided over this issue, and the conflict became translated into ideological terms by most club members. The supporters of Mr. Hi saw him as a fatherly figure who was their spiritual and physical mentor, and who was only trying to meet his own physical needs after seeing to theirs. The supporters of John A. and the other officers saw Mr. Hi as a paid employee who was trying to coerce his way into a higher salary. This situation ended by a fission of the club. The $34$ observed people of this dataset had different instructors at the beginning, and decided after the fission to move to either Mr. Hi's new club, or remain in the previous club.\\

\begin{figure}
\centering
\includegraphics[trim=0cm 0cm 0cm 0cm, clip=true, scale = 0.40]{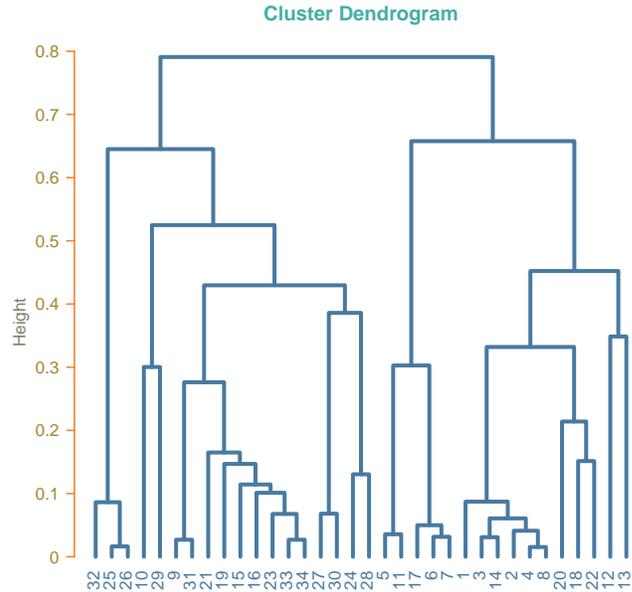}
\caption{\label{fig:Zach} Community estimation in Zachary's Karate club. Example of partitioning result.}
\end{figure}
A result of our algorithm is plotted in Fig. \ref{fig:Zach} as a Dendrogram. The corresponding $\mathcal{R}'$ is displayed in Table \ref{tble:zach}, with the optimal modularity value. $MS$ stands for \emph{Multiple Splitting}, which is the first algorithm we introduce. $BS$ stands for \emph{Binary Splitting} and refers to the algorithm presented in Section \ref{sec:Af}. Results of different other algorithms are also displayed showing better performances for some ($RW$, $SBM$), worst for others ($Betweeness$).\\
\begin{table}[t!]
\caption{\label{tble:zach}. {\small Comparison of the classification techniques for community detection. This table provides the Rand index and the maximum modularity value for each technique applied to the Zachary's Karate Club dataset.}}
\begin{ruledtabular}
\begin{tabular}{cccccc}
  & DS & MS & Betweenness \footnote{\cite{Newman2003a}} & RW \footnote{\cite{Pons2005}} & SBM \footnote{\cite{Newman2006}}\\
  \hline
  \hline
  $\mathcal{R}'$     &  $0.483$  & $0.509$   & $0.391$ & $0.590$ & $0.882$ \\

   Modularity     & $0.449$   & $0.459$  &  $0.401$ & $0.394$  & $0.382$ \\
\end{tabular}
\end{ruledtabular}
\end{table}
In this network, we are expected to find $2$ communities, representing where the people registered after the splitting of the club. This is however an empirical expectation, driven by the observations after the split, and there may not exist an optimal ground truth of communities. We can notice though that our community detection identified $4$ communities, hence the $\mathcal{R}'$ values. However, a cutting the tree in order to have two communities would lead to a value of $0.882$, identical to the best performance of a community detection algorithm.

\subsection{Political Books}

The data where compiled by \cite{Krebs}, in an unpublished work. The nodes represent books about US politics sold by the online bookseller Amazon.com. The edges represent frequent co-purchasing of books by the same buyers, as indicated by the "customers who bought this book also bought these other books" feature on Amazon. Nodes have been given values "l", "n", or "c" to indicate whether they are "liberal", "neutral", or "conservative".  These alignments were assigned separately by Mark Newman based on a reading of the descriptions and reviews of the books posted on Amazon.\\
\begin{figure}
\includegraphics[trim=0cm 2cm 0cm 2cm, clip=true, scale = 0.45]{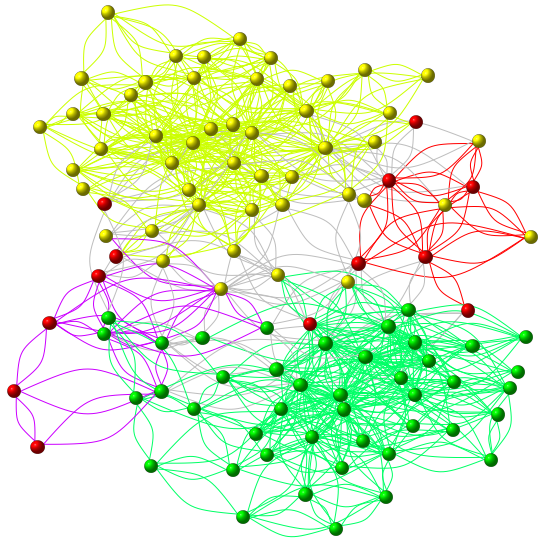}
\caption{\label{fig:books} Community estimation in political books. Example of community detection in the Political Books data.}
\end{figure}
A result of our algorithm is plotted in Fig. \ref{fig:books}. The coloured edges refer to the links within an estimated community, while the grey edges refer to links between different (estimated) communities. The ground truth is specified by the color of the spheres.
\begin{table}[t!]
\caption{\label{tble:polbooks}. {\small Comparison of the classification techniques for community detection. This table provides the Rand index and the maximum modularity value for each technique applied to the Political Books dataset.}}
\begin{ruledtabular}
\begin{tabular}{cccccc}
  & DS & MS & Betweenness\footnote{\cite{Newman2003a}} & RW\footnote{\cite{Pons2005}} & SBM\footnote{\cite{Newman2006}}\\
  \hline
  \hline
  $\mathcal{R}'$     &  $0.528$  & $0.519$   & $0.682$ & $0.653$ & $0.282$ \\

   Modularity     & $0.541$   & $0.522$  &  $0.517$ & $0.507$  & $0.391$ \\
\end{tabular}
\end{ruledtabular}
\end{table}
The results of the algorithms are displayed in Table \ref{tble:polbooks}. The number of communities is slightly overestimated ($4$ and $5$), leading to a value of $\mathcal{R}'$ equal $0.528$ and $0.519$. In some sense, the nature of the data makes it understandable to confuse the "neutral" with the "conservative" or "liberal". And that's mainly what we can observe from figure \ref{fig:books}. In fact, only three books were confused between "liberal" and "conservative", out of $92$. Once again, the knowledge of the number of communities will see the $\mathcal{R}'$ value raise to $0.688$, matching the best performance of the proposed algorithm in the literature.

\section{Conclusion}

We presented in this paper an algorithm to detect communities in networks, building on existing techniques and introducing new approaches. The aim was to simplify the process of detecting communities by reducing the dimension of the problem and introducing a more complete model taking into account the way the information is spreading within a network.\\
As demonstrated in the application examples, the results are more robust than the ones provided by the other leading techniques, as can be seen from Table \ref{tble:zach} and \ref{tble:polbooks}. Moreover, our approach is easily scalable to higher dimensional networks at lower costs, and is very easy to interpret, when more complex models fail at this feature.

\bibliography{./Publications-SMC_Com_Det.bib}

\end{document}